\documentclass[12pt]{article}
\usepackage{geometry}                
\usepackage{graphicx,amssymb}
\usepackage{rotating}
\usepackage{dcolumn}
\usepackage{amsfonts}
\usepackage{amssymb}

\usepackage{amsmath}
\usepackage{latexsym}
\usepackage{epsfig}
\usepackage{dsfont,eucal,mathrsfs}
\usepackage{verbatim}
\usepackage[textsize=footnotesize]{todonotes}
\usepackage{color}
\usepackage{amsfonts}
\usepackage{amsmath}
\usepackage{amsthm}
\usepackage{bm}

\newcommand{\keywords}[1]{\par\addvspace\baselineskip\noindent\textbf{Keywords:}\enspace\ignorespaces#1}

\begin{document}

\title{Fast Fermi Acceleration and Entropy Growth}

\author{Tiago Pereira$^{1,2}$ and Dmitry Turaev$^{1,3}$ \\
{\small $^1$ Department of Mathematics, Imperial College London, SW7 2AZ, UK} \\
{\small $^2$ Instituto de Ci\^encias Matem\'aticas e de Computa\c{c}\~ao}  \\
{\small Universidade de S\~ao Paulo, S\~ao Carlos, Brazil,} \\
{\small $^3$ Lobachevsky University of Nizhny Novgorod, 603950 Russia}}

\date{}

\maketitle

\abstract{Fermi acceleration is the process of energy transfer from massive objects in slow motion to light objects that move fast. The model for such process is a time-dependent Hamiltonian system. As the parameters of the system change with time, the energy is no longer conserved, which makes the acceleration possible. One of the main problems is how to generate a sustained and robust energy growth. We show that the non-ergodicity of any chaotic Hamiltonian system must universally lead to the exponential growth of energy at a slow periodic variation of parameters. We build a model for this process in terms of a Geometric Brownian Motion with a positive drift and relate it to the entropy increase.}

%*******************************************************************
%KEYWORDS
%*******************************************************************

% these are examples, put your key words

%*******************************************************************
%AMS SUBJECT CLASSIFICATION
%*******************************************************************
%\subjclass{35Q53\sep 34B20\sep 35G31}

% these are examples, put your numbers

%*******************************************************************
%ARTICLE BODY
%*******************************************************************
%\titlerunning{Fast Fermi Acceleration}

\maketitle

%*******************************************************************
%DO NOT FORGET TO RESET THE EQUATION COUNTER TO 0 AT THE HEAD OF EACH SECTION
%*******************************************************************

\section{Introduction}

Consider a family of Hamiltonians $H(p,q;\tau)$, where $\tau$ denotes a set of parameters changing slowly with time.
The equations of motion are
\begin{equation}\label{sys1}
\dot q= \frac{\partial H}{\partial p} (p,q;\tau), \qquad \dot p= - \frac{\partial H}{\partial p} (p,q;\tau).
\end{equation}
As parameters change, the energy $E=H(p,q;\tau)$ is not preserved by the system: 
\begin{equation}\label{eneq}
\dot E =\frac{\partial H}{\partial\tau} \dot\tau.
\end{equation}
Our goal in this paper is twofold: we  show that under very general conditions periodic oscillations of parameters lead to the exponential energy increase, and
we investigate the main features of this process.

Physically, model (\ref{sys1}) accounts for the energy transfer from massive objects in slow motion to light objects performing a fast motion.
The $(p,q)$ variables are momenta and coordinates that correspond to the fast degrees of freedom, while the parameters $\tau$ describe the effect
of the slow degrees of freedom (we assume that they correspond to a certain massive object, i.e., their evolution is not influenced by the fast dynamics).
It is known since the Fermi's work on cosmic rays \cite{Fermi} that acceleration is possible in this setting, see also \cite{Ulam}. In addition to
cosmic rays studies \cite{Kochar}, the interest to the problem is also motivated by plasma confinement \cite{plasma} and nuclear fission \cite{nuc}.

Previous research has been most often focused on time-dependent billiard models. Billiard is a dynamical system corresponding to a particle which moves
inertially inside a bounded domain and, upon hitting the boundary of the domain, reflects elastically. In time-dependent billiards, the boundary moves
according to a certain given law which is not affected by the collisions with the particle (i.e., the billiard wall is infinitely heavy). The wall motion changes
the particle reflection law, so the particle's kinetic energy is no longer conserved at the collisions. Still, this does not lead to an acceleration
in the one-dimensional case (where the particle moves inside an interval with slowly and periodically moving end points). Here, the kinetic energy
stays bounded for all times \cite{1dpeople,P95,LL,Dov}, provided the law of the end points' motion is smooth enough. As was discovered in \cite{LRA1,LRA2}, in
the two-dimensional case the situation changes. When the domain's boundary slowly moves so that the dynamics inside the corresponding frozen billiard
is chaotic the time-dependent billiard produces accelerating particles, as can be seen from various numerical experiments \cite{2dpeople,CSL1,CSL2,LDS,LOL,OVL,BR}.
The theory of this phenomenon was even earlier proposed by Jarzynski who derived a universal model describing the evolution of the distribution of
energies in a strongly chaotic billiard with moving walls \cite{Jar}. In these examples, the observed acceleration was relatively slow,
giving at most polynomial in time growth of the energy averaged over a uniform ensemble of initial condition.
Such regime can be easily destroyed by a small dissipation \cite{BunimovichLeonel,OR}.

It was shown in \cite{GRST,GRT} that the main limitation for fast energy growth is imposed by the existence of an Anosov-Kasuga adiabatic
invariant \cite{An,Kas,Ott,Gre,McK}. This adiabatic invariant is a function of the particle energy and the slow time $\tau$. It follows from the
Anosov averaging theorem \cite{An,LoM} that the change of its logarithm over the period of the billiard wall oscillations tends to zero for the majority of initial conditions
as the particle energy grows. This means that the rate of possible acceleration must slow down with the energy increase.
The logarithm of Anosov-Kasuga invariant coincides with the Hertz entropy of the frozen billiard \cite{Herzentr}. Therefore, its approximate preservation is in
full agreement with the basic physical intuition: the entropy of an ergodic system does not grow at slow (i.e., adiabatic) changes of system's parameters.

Recently, various examples of time-dependent billiards with {\em exponential} energy growth were
produced \cite{RST,GRST,Kushal,mushroom,bati}. In these works, the billiard wall moves in such a way that the
frozen billiard loses its ergodicity (i.e., it has more than one ergodic component) at least at the part of the period of the boundary oscillations.
These works ergodicity violation, but producing the same effect: The Anosov averaging theorem does not hold
for non-ergodic systems, so the Anosov-Kasuga invariant no longer exists, and the energy grows unimpeded with each period of the billiard wall oscillations.

In this paper we depart from the billiard setting, and investigate a general question: How can an adiabatic periodic variation of parameters lead to a sustained energy growth? The Anosov-Kasuga adiabatic invariant
is not billiard specific, so it imposes restrictions to the energy growth in any Hamiltonian system with slowly
changing parameters if the frozen dynamics is ergodic on every energy level. However, apart from special classes of systems,
such as geodesic flows and billiards, {\em a typical Hamiltonian system is not ergodic}.

We demonstrate that the non-ergodicity of a chaotic Hamiltonian system must universally lead to the exponential growth of energy at a slow periodic oscillation of
parameters. The key mechanism is the following: A non-ergodic Hamiltonian system has regions of chaotic dynamics in the phase space, which coexist with
stability islands where dynamics is nearly integrable (quasiperiodic). Adiabatic changes of parameters lead to transitions between these regions. Different
initial conditions give rise to different itineraries of these transitions, and different itineraries give different values of the energy gain/loss
per period of the parameters oscillation. We introduce an analogue of Hertz entropy for non-ergodic systems and show that on average over all possible itineraries
the entropy linearly increases after each period. This yields the exponential energy growth.

\subsection{Assumptions and claims}\label{ac}

Consider a family of Hamiltonians $H(p,q;\tau)$ where $q,p \in \mathbb{R}^n$ and $\tau$ is a set of parameters. We assume that
\begin{itemize}
\item[A1] the Hamiltonians in the family are homo\-geneous at each frozen $\tau$,
\item[A2] the parameters $\tau$ change periodically with time,
\item[A3] the Hamiltonian $H(p,q;\tau)$ has more than one ergodic component (on each energy level) at each frozen value of $\tau$ for at least part of the period.
\end{itemize}

Our hypothesis $A1$ means that each Hamiltonian is invariant with respect to energy scaling, so the dynamics on each energy
level is the same. A typical example is the motion in a homogeneous polynomial potential, see e.g. (\ref{hv4}). Another
example is the Boltzmann gas of hard spheres. Note that assumption $A1$ is not restrictive. We may start with an arbitrary system, but
since we study the process of an unbounded energy growth, we must consider the system in the high-energy limit, and it is most natural
to assume that the system becomes invariant with respect to the energy scaling in this limit. For example, if we start with a general
polynomial potential, only the highest order terms will be most relevant at high energies, i.e., the potential will effectively become homogeneous.
For the motion in a potential confined to a bounded domain $D$ (i.e., the potential which is finite inside $D$ and infinite on the boundary of $D$)
the high-energy limit is the billiard in $D$ \cite{TRK}, which is also a homogeneous system. In essence, our homogeneity assumption means that the external
forcing is applied at the highest order terms, i.e., the forcing is of the same order as the energy:
\begin{equation}\label{hth}
\frac{\partial H}{\partial \tau} \sim H.
\end{equation}

Unless we consider relativistic particles, the values of $\dot q$ and $\dot p$ grow with the energy while $\tau$ remains bounded.
Thus, at large energies, the variables $(p,q)$ are fast and the parameters $\tau$ are varying slowly. Another slow variable is
the energy $E=H(p(t),q(t);\tau(t))$: since $\partial H/\partial \tau \sim H$, the speed of change of $\ln E$ is comparable with $\dot\tau$.

A proper way to understand the evolution of slow variables is to somehow average the system over the fast variables. Such model reduction
goes naturally when the frozen system is ergodic with respect to the Liouville measure, i.e., Lebesgue measure restricted to the constant energy level:
$$
\mu_L(E,\tau; dp dq)=\delta(E-H(p,q;\tau)) dp dq.
$$
In this case Anosov averaging theorem \cite{An,LoM} guarantees that the evolution of the energy can be described by
averaging equation (\ref{eneq}) over $\mu_L$, that is
$$
\displaystyle \dot E= \left( \frac{\int \frac{\partial H}{\partial \tau}(p,q;\tau)\mu_L(E,\tau; dp dq)}
{ \int \mu_L(E,\tau; dp dq)}\right)  \dot\tau,
$$
with good accuracy and for a large set of initial conditions, see details in Section \ref{entropy}. In the non-ergodic case there is no unique measure
the averaging over which could provide a description of the slow evolution of energy which would be valid for the majority of initial conditions simultaneously.
Indeed, below we propose a model where the energy change is governed by equations obtained
by averaging over certain {\em randomly chosen} ergodic components, exact choice of which depends on the initial conditions.

The main approximation principle we use in our approach is that we assume that
{\em at every value of the slow variables the motion of the fast variables immediately ergodises}.
One can think of this as at each value of $\tau$ and $E$ the phase point is uniformly distributed over a certain ergodic component of the fast system.
We stress that we view this only as an approximation, i.e., our assumption is that the time averages
over intervals long in terms of the fast variables and short in terms of the slow variables coincide with averages over a certain ergodic
component of the fast system with a sufficiently good accuracy for a sufficiently large set of initial conditions.

In this approximation, at each value of $\tau$ and $E$ the evolution of $E$ is defined by
\begin{equation}\label{even}
\dot E = \left\langle \frac{\partial H}{\partial \tau} \right\rangle\; \dot\tau
\end{equation}
where the averaging is done over a certain ergodic invariant measure of the frozen system on the corresponding energy level. Moreover, the fast ergodisation assumption
also implies that the probability to switch in this equation from one ergodic component to another (as $\tau$ changes) depends only on the pair of ergodic components
involved. Thus, we have a time-dependent (periodic) Markov process on the set of ergodic components of the frozen system. Without losing much precision
one can discretise time and also approximate the set of ergodic invariant measures by a finite set, so one obtains a process of hopping over a finite set of states.
Different initial conditions correspond to different realisations of this process, i.e., to different time-dependent families of ergodic invariant measures;
we call such a family {\em an averaging protocol}. Each protocol produces its own averaged equation for the evolution of energy.

Similar construction was described in \cite{bati,TV}. The validity of such approach was numerically checked in \cite{GRST,mushroom,bati} for various
examples of time-dependent billiards. In these examples, the structure of ergodic components was known, so the transition probabilities for the hopping
process were computable. In a general situation (e.g. in the example we consider in Section \ref{numerics})
there is no hope of knowing the structure of the set of ergodic components, nor computing the transition probabilities.
Therefore, we formulate the main properties of the resulting model, which are independent of its particular details.

\bigskip

{\bf Multiplicative law for the energy evolution.}
The first conclusion we make is that after each period $T$ of the parameters' oscillations the energy is multiplied to a random factor:
\begin{equation}\label{claim1}
E_{n+1} = E_n \zeta_n,
\end{equation}
where $E_n = H(p,q;\tau(nT))$ is the energy after $n$ periods $T$. The multiplicative character of law (\ref{claim1}) is due to
our assumption (\ref{hth}) which implies that $\dot E\sim E$ in (\ref{even}) no matter which ergodic component is chosen to average over.
The energy gain factor $\zeta_n$ depends on the particular averaging protocol. As the sequence $\zeta_n$ is formed via a Markov hopping process which typically
exhibits an exponential decay of correlations, the correlations between consecutive values of $\zeta_n$ must rapidly decay. For simplicity we just assume
that the factors $\zeta_n$ are independent random variables; they are also identically distributed (as the system we consider depends on time periodically).

{\bf Exponential growth of energy.} Model (\ref{claim1}) describes a random walk for $\ln E_n$. By the Law of Large Numbers,
for a typical realisation of the random walk (i.e., for a typical initial condition) we must have
$$\lim_{n\to+\infty} \frac{1}{n}\ln E_n=\mathbb{E} \ln \zeta.$$
In Section \ref{entropy} we introduce a notion of entropy for our system and show that the preservation of volumes in the $(p,q)$ space
by system (\ref{sys1}) implies that the random process (\ref{claim1}) cannot decrease the entropy. In the ergodic case the entropy would
be preserved, but in the general case of an adiabatically perturbed homogeneous Hamiltonian system with several distinct ergodic components
in each energy level we expect that {\em the entropy increases to a non-zero quantity with each period} $T$.

This is equivalent to the claim that the multiplicative random walk model (\ref{claim1}) has a positive bias:
\begin{equation}\label{claim2}
\rho=\mathbb{E} \ln \zeta > 0.
\end{equation}
Hence,  {\it the energy grows exponentially both for typical initial conditions and on average}.
As we mentioned, the non-ergodicity plays an important role here: In the ergodic case the bias $\rho$ vanishes and model (\ref{claim1}) becomes
invalid (the energy grows at most polynomially \cite{GRST,GRT}).

Note that since the factors $\zeta_n$ in (\ref{claim1}) are assumed to be independent, the energy averaged over all initial conditions grows by the law
$$\mathbb{E} E_{n+1}= (\mathbb{E} \zeta_n) \;\mathbb{E} E_n.$$
This corresponds to the exponential growth with the rate
$$\frac{1}{n}\ln \mathbb{E} E_n = \rho^+ = \ln \mathbb{E} \zeta.$$
As $\rho^+>\rho$, the averaged energy growth is faster than the energy growth for a typical conditions,
which means that a small minority of realisations far outperform the rest.

{\bf Log-Gaussian distribution.} The positivity of the rate $\rho$ is the universal characteristics of the energy growth
process in adiabatically perturbed non-ergodic systems. The distribution of energies formed at this process
is not universal, as billiard examples of \cite{GRST,mushroom,bati} show. As $\ln E_n$ is, according to (\ref{claim1}), the sum
of independent bounded random variables, the logarithm of the energy becomes, after a proper scaling, normally distributed
at large $n$. This does not mean that the energy itself acquires a Log-Gaussian distribution (since the scaling and taking a logarithm
do not commute). However, in the example we consider in this paper the energy distribution is, in fact, close to Log-Gaussian, right after the first period
(see Figs. \ref{Fig_EnD},\ref{Fig_check}).

\begin{figure}[!ht]
\centering
\includegraphics[width=10cm]{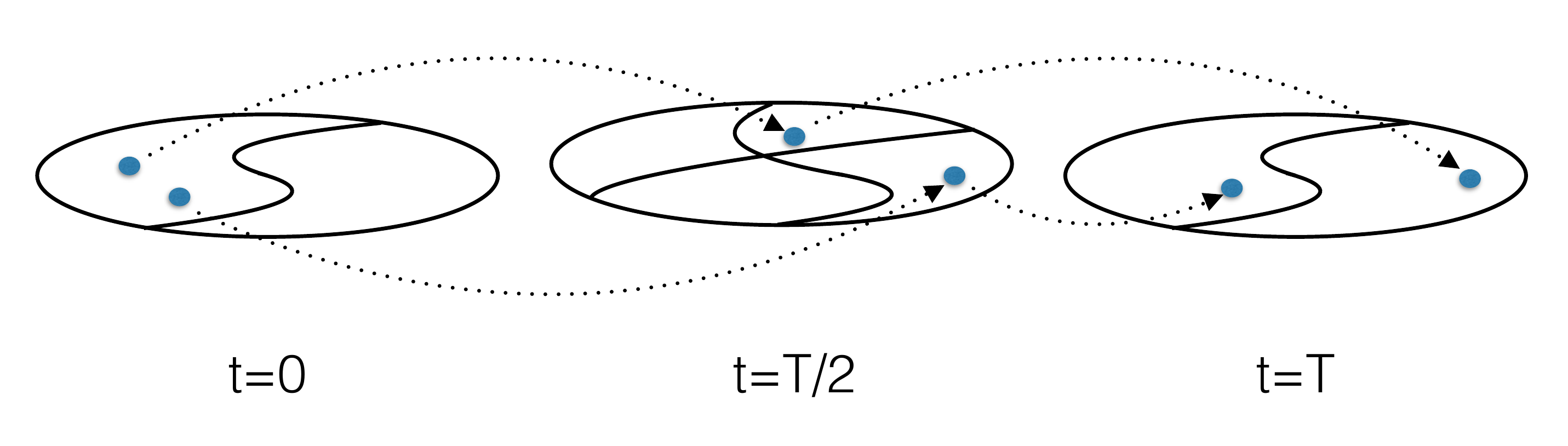}
\caption{Illustration of the jumping process between macro-states. Initial points will have different itineraries visiting distinct ergodic components during the period.}
\label{jumping}
\end{figure}

We believe that this must be typical for the case where the frozen Hamiltonian system has many ergodic components (at least for a part of the period $T$),
so a typical trajectory will hop many times between the components during the period, see Figure \ref{jumping} for an illustration. By (\ref{even}) and (\ref{hth}), if for a certain time interval the
change of energy along the orbit is determined by a given ergodic component, the energy at the end of this interval equals to the energy at the beginning
of the interval times a certain factor; the product of these factors over the period gives the multiplier $\zeta$ from (\ref{claim1}). Thus, if the number of jumps
between the components is large, then $\zeta$ occurs to be a product of a large number of random factors, each of which must be close to $1$ (as $\zeta$ must
remain bounded). This means that the logarithmic gain $\ln \zeta$ at the end of the period is in the Central Limit Theorem regime, i.e.,
the distribution of $\ln\zeta$ is close to a Gaussian. Since the sum of independent
Gaussian variables is also Gaussian, we conclude that the distribution of the scaled energy $E_n e^{-n\rho}$ after $n$ periods is close to Log-Gaussian; moreover,
we can estimate parameters of this distribution:
\begin{equation}\label{distLogE}
\nu (\ln E_n) \approx \mathcal N (n \rho, \sqrt{n}\sigma)
\end{equation}
where $\mathcal N (a, b)$ denotes the normal distribution with the mean $a$ and standard deviation $b$, and
\begin{equation}\label{rsrrp}
\sigma^2=\mathbb{E} (\ln \zeta - \rho)^2 = 2(\rho^+-\rho).
\end{equation}
We note that except for specially prepared examples (like above mentioned billiards) it is impossible to theoretically compute the rates $\rho$ and $\rho^+$;
however, we believe that relation (\ref{rsrrp}) between the energy growth rates and the width of the distribution of the logarithmic gain must hold
for a large class of (not specially prepared) systems.

This particular class of multiplicative random processes, where the logarithmic gain is Gaussian, is called geometric Brownian motion (GBM).
It emerges in various applications, notably in finance, and is well studied, see e.g. \cite{Ok,Ole} and references therein.

\section{Entropy growth}\label{entropy}

\begin{figure}[!ht]
\centering
\includegraphics[width=5cm]{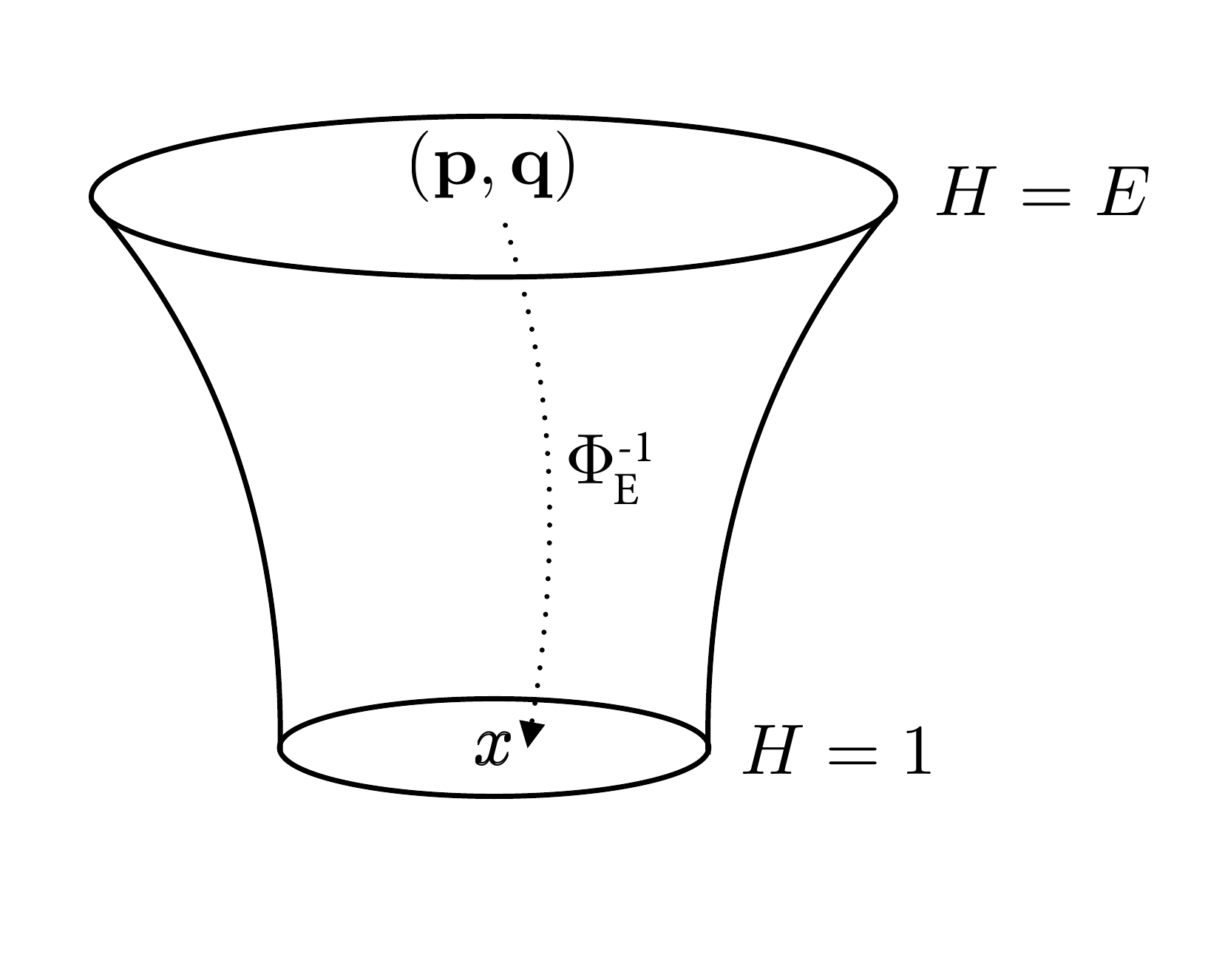}
\caption{Illustration to the $(x,E)$-coordinates.}
\label{Illustration1}
\end{figure}

Now we give a more detailed description of the model, introduce the notion of entropy, and demonstrate the validity of (\ref{claim2}).
We call a Hamiltonian $H(p,q)$ {\em homogeneous} if for any $E>0$ there exists a coordinate transformation $\Phi_E$ that keeps
the system the same, sends the energy level $H=1$ to $H=E$, and has a constant Jacobian
$$
J(E)=E^\alpha, \, \, \,  \alpha>0.
$$
We assume that the positive energy levels are compact, so
\begin{equation}\label{jve}
V(E)=V(1) J(E)= V(1) E^\alpha
\end{equation}
where $V(E)$ is the volume of the $(p,q)$-space between the energy levels $H=E$ and $H=0$.
Thus, we can label the points in the phase space $(p,q)$ by the coordinates $(x,E)$ where $E=H(p,q)$ is the energy and $x=\Phi_E^{-1}(p,q)$ is the projection
to the energy level $H=1$, see Fig. \ref{Illustration1}.

We consider a $\tau$-dependent family of homogeneous Hamiltonians $H(p,q;\tau)$, so the transformation $\Phi$ and the volume $V$ are now also functions of $\tau$.
We assume that $\displaystyle \frac{\partial}{\partial \tau} \ln H= \left(\frac{\partial H}{\partial\tau}\right)/H$
does not depend on energy.This means that $\frac{\partial H}{\partial\tau}$ scales in the same way as $H$ and condition (\ref{hth}) holds.
We also introduce the notation
\begin{equation}\label{drlh}
G(x,\tau):= \frac{\partial}{\partial \tau} \ln H(q,p;\tau).
\end{equation}

Let us now allow the parameters to change periodically with time. We assume that at large energies the motion in the frozen
system (that corresponds to a fixed value of $\tau$) is fast, i.e., the variables $x$ change, at large energies, much faster
than $\tau$ does. By (\ref{hth}), the speed of change of $\ln E$ is
comparable with $\dot\tau$. Thus, we have a slow-fast system, with fast variables $x$ and slow variables $\tau$ and $\ln E$.

If the frozen system is {\em ergodic} on any energy level with respect to
the Liouville measure $\mu_L$, we can apply Anosov theorem \cite{An}. This theorem guarantees, that for any given finite number of periods
and any given accuracy, if the initial value of energy is taken large enough, then the evolution of the logarithm of energy
is, within this accuracy, described on the chosen time interval by the averaged equation
\begin{equation}\label{ergc}
\frac{d}{dt} \ln E=\frac{\displaystyle \int \frac{\partial H}{\partial\tau}(p,q,\tau) \delta(E-H(p,q,\tau)) dpdq}
{\displaystyle E \int \delta(E-H(p,q,\tau)) dpdq} \; \dot\tau
\end{equation}
for {\em the majority of initial conditions} (i.e., for all initial conditions except for a set of small measure). By changing variables $(p,q)$ to $(E,x=\Phi_E^{-1}(p,q))$,
we reduce this equation to
\begin{equation}\label{ergcx}
\frac{d}{dt} \ln E=\frac{1}{\int dx}  \left( \int G(x,\tau) dx \right)  \; \dot\tau,
\end{equation}
where $G$ is defined by (\ref{drlh}); the variable $x$ runs the level $H=1$.

We note that no mixing is required, i.e., one should not think of a fast relaxation to the Liouville measure at every
moment of the slow time. Just the measure of the set of initial conditions for which the evolution of energy deviates
noticeably from that given by this averaged equation is small (the measure of the bad set of initial conditions can be made as small as we want by taking
the initial energy value large enough).

Note that equation (\ref{ergc}) implies
$$
\frac{d}{dt}{V}(E,\tau)=\frac{\partial V}{\partial E} \dot E + \frac{\partial V}{\partial \tau} \dot \tau=0.
$$
Indeed, the volume under the energy level $H=E$ is given by
$$V(E,\tau)=\int_{H(p,q,\tau)\leq E} dpdq=\int \theta(E-H(p,q,\tau)) dpdq$$
where $\theta$ is the Heaviside function. Since $\theta'=\delta$, it follows that
$$\frac{\partial V}{\partial \tau}=-\int \frac{\partial H}{\partial\tau}(p,q,\tau) \delta(E-H(p,q,\tau)) dpdq,$$
$$\frac{\partial V}{\partial \tau}=\int \delta(E-H(p,q,\tau)) dpdq,$$
so
$$\frac{d}{dt} V(E,\tau)=0$$
by (\ref{ergc}) (cf. \cite{Kas,Ott,Gre,McK,Herzentr}). This means that energy changes periodically with $\tau$ (to keep $V(E,\tau)$ constant).
As the behaviour of $\ln E$ in the original system is close to that given by (\ref{ergc}), we conclude, using (\ref{jve}), that
for the large majority of initial conditions $\ln V(E,\tau)$ stays close to its initial value for the given in advance number
of periods, $\ln E$ stays bounded and, at the end of each period, returns close to its initial value $\ln E_0$, i.e.,
$E/E_0$ returns close to $1$ at the end of each period. This shows that if the frozen system is ergodic, then
there can be no significant energy growth (nor decrease) for a long interval of time and a large set of initial conditions.

\bigskip

In the {\em non-erdodic} case this conclusion is no longer valid. Moreover, it is no longer true that the evolution of
energy is approximately the same for the majority of initial conditions. Maximum we can hope, in analogy with the Anosov averaging theorem,
that the slow evolution of the energy is described (for a given number of slow periods, with a prescribed accuracy, and for a large set of initial condition)
by the following equation (cf. (\ref{ergcx})):
\begin{equation}\label{mna}
\frac{d}{dt} \ln E= \left(\int G(x,\tau) \mu_{\tau}(dx) \right) \; \dot\tau
\end{equation}
where $\mu_\tau$ is, at each value of $\tau$, a certain ergodic measure (invariant with respect to the frozen system) on the energy level $H=1$.
Moreover, as we are looking only for a finite accuracy approximation, we can replace the measures $\mu_\tau$ by one of a finite set of measures
$$\mu_{i\tau}=\xi_i(\tau,x) \frac{dx}{\int_{H=1} dx}$$
where $\xi_i$, at every given $\tau$, are characteristic functions of certain open sets which form a partition of the $x$-space; we also may
assume that the dependence of $\xi_i$ on $\tau$ is piece-wise constant.

We call the measures $\mu_i$ (or their support sets) {\em macro-states}. The orbit of the system hops, as $\tau$ changes, between different macro-states. As we mentioned
in the previous Section, a part of our fast ergodisation principle is that the corresponding sequences of macro-states (the averaging protocols)
are realisations of a $\tau$-dependent Markov process. For a typical Markov process, for any given distribution of initial conditions over the macro-states,
after a few periods of oscillation of $\tau$ the distribution will converge to a certain time-periodic distribution. Thus, independently of the given
initial density of phase points in the $(x,E)$-space, we may assume that after a few periods $T$ the dynamics will form the same piecewise constant density in the $x$-space
at the beginning and the end of each period; this equilibrium density will take constant values on macro-states.

The next question is to investigate the dynamics of the distribution of energies. Note that in our construction we have a finite set of averaging protocols
over one period $T$ of oscillations of $\tau$ (the number $K$ of the protocols depends
on the accuracy of the approximation we want to achieve, but it does not depends on the energy, i.e., our model provides the same accuracy of approximation
on a given number of periods for all sufficiently large values of energy). Let us split the space of initial conditions $x$ into
{\em meso-cells} $M_1, \dots, M_K$ that give rise to the distinct averaging protocols. By construction, each cell is a subset of one of the macro-states
that exist at the beginning of the period, and the image of the cell by the flow of the system after the period $T$ is also a subset of one of these macro-states.
For each cell the majority of initial conditions gives rise to the same energy evolution (given by equation (\ref{mna})) over the period $T$,
while for initial conditions from different cells the values of energy gain or loss will be different.

Let $E_0$ and $E_1$ be two sufficiently large values of energy. Choose a given cell $M_k$ ($k=1,\dots,K$). By equation (\ref{mna}),
if the orbits with initial conditions $E=E_0$, $x\in M_k$ move to the energy  level $E=\bar E_0=e^{\lambda_k} E_0$
after the period $T$, then likewise, the orbits with initial conditions $E=E_1$, $x\in M_k$
move to the level $E=\bar E_1=e^{\lambda_k} E_1$. Now note that the non-averaged system preserves volume in the $(p,q)$-space.
Therefore, it follows that the volume occupied by the points with $x\in M_k$ between the levels $E=E_0$ and $E=E_1$ equals to the volume occupied
by the points with $x\in \bar M_k$ between the levels $E=\bar E_0$ and $E=\bar E_1$, where $\bar M_k$ denotes the image of the cell $M_k$
by the flow of the (non-averaged) system after the period $T$. This gives
$$v(M_k)(V(E_1)-V(E_0))=v(\bar M_k) (V(\bar E_1)-V(\bar E_0))$$
where $v$ is the volume in the $x$-space (the level $H=1$), and $V(E)$ is the volume under the energy level $H=E$ (all volumes here are computed
at the beginning of the period $T$, or at the end, which is the same). By plugging (\ref{jve}) into this formula,
we obtain the following relation between the energy growth rate $\lambda_k$ at the cell $M_k$ and the volumes of the cell $M_k$ and its image $\bar M_k$
after one period:
\begin{equation}\label{vlal}
\alpha\lambda_k=\ln \frac{v(M_k)}{v(\bar M_k)}.
\end{equation}
We also normalise $v$ so that the total volume of the $x$-space at the beginning of the period equals to $1$, that is,
\begin{equation}\label{vkbvk}
\sum v(M_k)=\sum v(\bar M_k) =1.
\end{equation}
~

Given a distribution $\nu$ of the points in the $(x,E)$-space, we define the corresponding {\em entropy} as
\begin{equation}\label{sdef}
S=\int \ln \frac{V(E,\tau)}{V(1,\tau)} \nu(dE,dx)=\alpha \int \ln E \;\nu(dE,dx),
\end{equation}
see (\ref{jve}). In our model, the energy undergoes the same evolution for the majority of the orbits that start at the cell $M_k$ ---
for all of them $\ln E$ gets an increment $\lambda_k$. Therefore, the change of the entropy over the period equals to
$$\Delta S= \alpha \sum_k \lambda_k \nu_k$$
where $\nu_k$ is the probability to be at the cell $M_k$ at the beginning of the period. By (\ref{vlal}), we obtain
\begin{equation}\label{dlts}
\Delta S= \sum_k \ln \left( \frac{v(M_k)}{v(\bar M_k)} \right) \nu_k.
\end{equation}

The result depends on the distribution $\nu$. However, as we mentioned, in our model any initial density in the $x$-space
converges to an equilibrium one, constant on macro-states. For example, there can be a situation where at some value of $\tau$
during the period the frozen system appears to be completely chaotic, i.e., at this value of $\tau$ there is only one macro-state,
hence, the corresponding equilibrium density is uniform. In this case, we choose this value of time to be the beginning of the period,
so $\nu_k = v(M_k)$ in (\ref{dlts}) (recall that $v$ is the scaled volume so that (\ref{vkbvk}) holds).
Note that we do not insist on ergodicity or mixing at this value of $\tau$. All our formulas
can be true only within certain, chosen in advance finite accuracy, so we just need that the uniform density would be close to our
equilibrium density with the given accuracy. This seems to be the situation we have in the example considered in Section \ref{numerics}:
the numerically obtained phase portrait at the beginning of the period seems fairly uniform.

Thus, if we have only one macro-state at the beginning of the period, then
$$
\Delta S= - \sum_k v_k \ln \frac{\bar v_k}{v_k}
$$
where $v_k=v(M_k)$, $\bar v_k=v(\bar M_k)$. As $\sum v_k=1$, and the geometric mean is smaller than the arithmetic mean,
we have
$$
\Delta S= - \sum_k v_k \ln \frac{\bar v_k}{v_k}\geq - \ln \left(\sum_k v_k \frac{\bar v_k}{v_k}\right)=-\ln \left(\sum_k \bar v_k\right)=0
$$
(see (\ref{vkbvk})). Note that $\Delta S$ can be zero only if $v_k=\bar v_k$ for all $k$. By (\ref{vlal}), this would mean no energy change for each averaging
protocol, which seems to be a quite degenerate phenomenon for the non-ergodic case. Therefore, unless we are in some degenerate situation,
{\em the entropy must acquire the same positive increment at each period}. By the definition of entropy (\ref{sdef}), the linear growth of entropy
means an exponential energy growth.

\bigskip

The same conclusion holds in the general case where we have more than one macro-state at the beginning of the period. In this case we have
a certain equilibrium density in the $x$-space, which produces the probabilities $\nu_k$ of being at the cell $M_k$, the same at the beginning
of each period. Recall that the equilibrium density is constant on a macro-state, and that each cell $M_k$
is a subset of a certain macro-state, so the equilibrium density is uniform on $M_k$. Therefore, the equilibrium density is given by
\begin{equation}\label{eqm}
\nu(x)=\nu_k/v(M_k)  \, \, \,  \mbox{   for   } \, \, \,  x \in M_k,
\end{equation}
where $v$ is the volume in the $x$-space. The points that start at a cell $M_j$ come, after time $T$, to the image cell $\bar M_j$. It is a part of
a certain macro-state, and our fast ergodisation
principle (see Section \ref{ac}) tells us that the probability of a point from $\bar M_j$ to be anywhere in this macro-state is the same within
the accuracy of our approximation. Therefore, the probability $Q_{kj}$ to start at the cell $M_j$ and, after time $T$, land at the cell $M_k$,
equals to $v(\bar M_j \cap M_k)/v(\bar M_j)$. If we denote $v_k=v(M_k)$, $\bar v_j=v(\bar{M}_j)$, we obtain
\begin{equation}\label{vrqr}
v_k=\sum_j Q_{kj}\bar v_j
\end{equation}
(as $v(M_k)=\sum_j v(\bar M_j \cap M_k)$). Recall that the probabilities $\nu_k$ are the same at the beginning of each period, so we also have
\begin{equation}\label{invpr}
\nu_k=\sum_j Q_{kj} \nu_j.
\end{equation}

By (\ref{dlts}), the change of entropy over the period $S$ is given by
\begin{equation}\label{engr}
\Delta S= \sum_k \nu_k \ln  \left[\frac{v_k}{\bar v_k} \right]=\Sigma-\bar\Sigma,
\end{equation}
where
$$
\Sigma=\sum_k \nu_k \ln  \left[\frac{v_k}{\nu_k}\right]=-\int \nu(x) \ln \nu(x) dx   \, \, \,  \mbox{  and  } \, \, \,
\bar \Sigma=\sum_j \nu_j \ln  \left[\frac{\bar v_j}{\nu_j}\right]=
-\int \bar\nu(x) \ln\bar\nu(x) dx,
$$
i.e., $\Sigma$ is the Shannon-Gibbs entropy of the equilibrium density (\ref{eqm}), and $\bar\Sigma$ is the Shannon-Gibbs entropy of the density
$\bar\nu(x)=\frac{\nu_j}{\bar v_j}$ at $x\in \bar M_j$. The density $\bar\nu$ is the image of $\nu$ by the slow evolution over the oscillation period,
just before it is replaced by $\nu$ due to the fast ergodisation. The ergodisation must increase the entropy, so $\Delta S\geq 0$.

Indeed, let $\gamma_{kj}=Q_{kj}\nu_j/\nu_k$. Since $\sum_j \gamma_{kj}=1$ (see (\ref{invpr})), and the geometric mean is less than arithmetic mean,
we have
$$\sum_j \gamma_{kj}\ln(\bar v_j/\nu_j) = \ln \left(\prod_j \left(\frac{\bar v_j}{\nu_j}\right)^{\gamma_{kj}}\right) \leq
 \ln \left(\sum_j \frac{\bar v_j}{\nu_j}\gamma_{kj}\right) = \ln \left(\frac{1}{\nu_k}\sum_j Q_{kj} \bar v_j \right),$$
which, by (\ref{vrqr}), gives
$$\sum_j \gamma_{kj}\ln(\bar v_j/\nu_j) \leq \ln (v_k/\nu_k),$$
$$\sum_j Q_{kj}\nu_j \ln(\bar v_j/\nu_j)\leq \nu_k\ln(v_k/\nu_k),$$
hence $\bar\Sigma\leq \Sigma$ (as $\sum_k Q_{kj}=1$).
Thus, the entropy is a non-decreasing function of time in the general case as well.
Note that the value of $\Delta S$ we compute here is proportional to $\mathbb{E} \ln \zeta$ where $\zeta$ is the energy gain factor from
(\ref{claim1}), i.e., we have established our claim (\ref{claim2}).

\bigskip

An intuitive explanation for the entropy growth at an adiabatic change of parameters of a non-ergodic system can be obtained as follows. Let us think of
an adiabatically changing Hamiltonian system as a gas of non-interacting particles (different particles correspond to different initial conditions). As there is
absolutely no interaction, there is no equilibrium distribution in energies. However, in the ergodic case one still recovers the conservation of entropy.
In the non-ergodic case, the particles are, at each value of the parameter $\tau$, in different macro-states which correspond to different ergodic measures
$\mu_{i\tau}$ over which the fast dynamics is averaged. Thus, our gas can be considered as a mixture of different phases or fractions; the adiabatic change of
parameters can lead to particles changing their phase, so the relative densities of each fraction in the gas can vary, and this naturally leads to the entropy growth.

\section{Numerical Simulations}\label{numerics}

\subsection{Quartic Potential}

\begin{figure}[!ht]
\centering
\includegraphics[width=8.5cm]{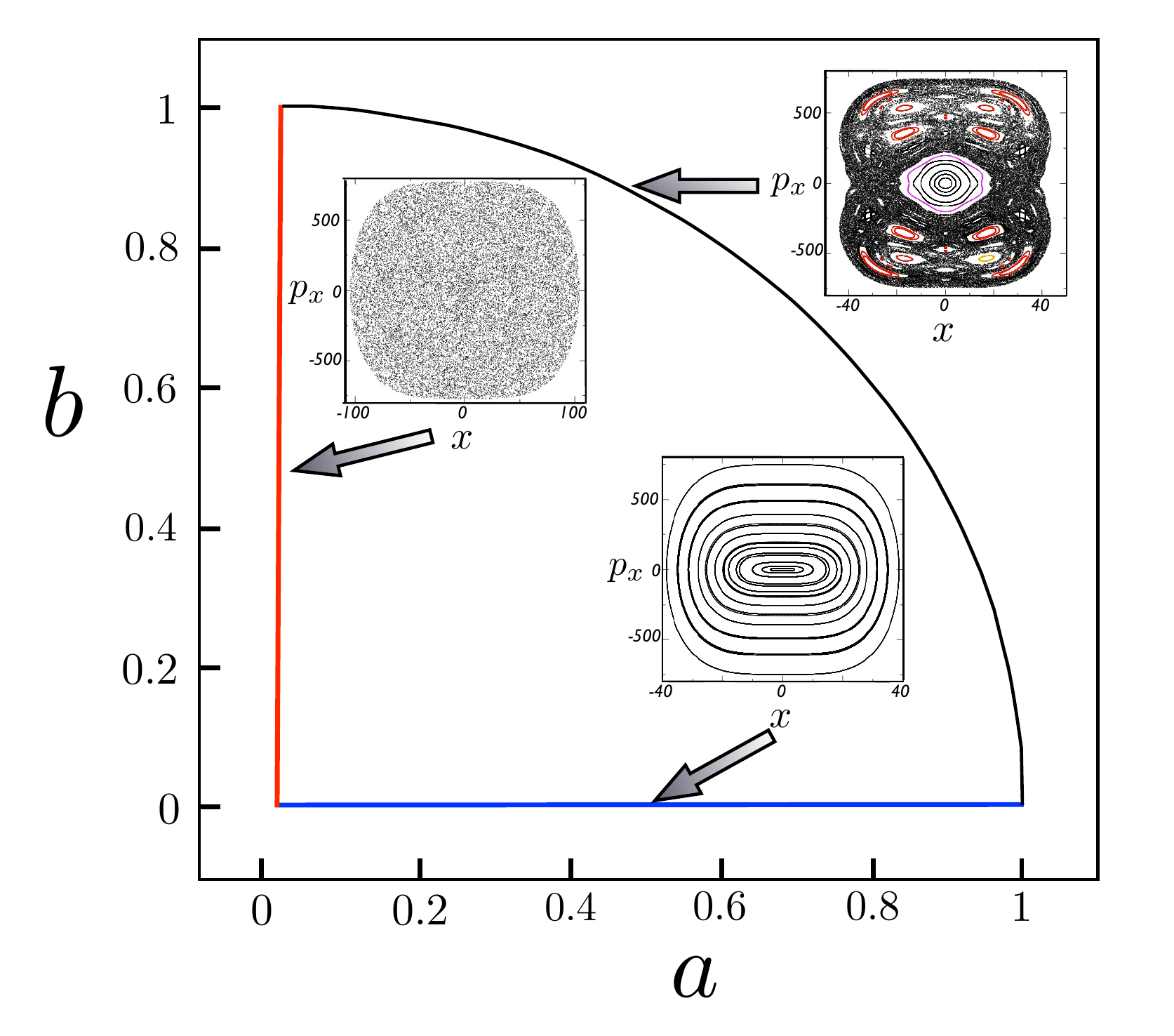}
\caption{{\bf Parameter plane $(a,b)$ for the quartic Hamiltonian and snapshots of dynamical behaviour}.
The insets show the dynamics of the Poincar\'e map on the cross-section $y=0$ (at $p_y>0$). Different
choices of the parameters $a$ and $b$ can lead to integrable and chaotic dynamics.}
\label{Fig_par}
\end{figure}

We verify our model in numerical experiments performed with
\begin{equation}\label{hv4}
H(\bm{p},\bm{q},\tau(t)) = \frac{p_1^2}{2}  + \frac{p_2^2}{2} + \frac{a(t)}{4}
\left( q_1^4 + q_2^4 \right)  + \frac{b(t)}{2} q_1^2 q_2^2,
\end{equation}
where $\bm{q} = (q_1,q_2)$, $\bm{p}=(p_1,p_2)$, and $\tau(t) = (a(t), b(t))$.
As the potential is homogeneous, the Hamiltonian has scaling properties, i.e., the dynamics of the frozen system
for any positive energy $E$ can be determined by simple scaling from the dynamics on the energy shell $E=1$.
For frozen values of the parameters, this system has been thoroughly studied \cite{Canergie,Bon}. The system is integrable in two situations:
\begin{enumerate}
\item For $a=b$ the conserved quantity as an angular momentum;
\item For $b=0$ the two degrees of freedom are uncoupled.
\end{enumerate}
There is also a chaotic regime; for example, for $a=0.01$, $b=1$ the system exhibits exponential decay of correlations.
Thus, we can change $a$ and $b$ in such a way that the system will undergo a transition between chaotic and integrable regimes,
see Fig. \ref{Fig_par}.

\begin{table}[ht]
\caption{Parameter Values used in the simulations.}
\centering
\begin{tabular}{ c c c} % Column formatting, @{} suppresses leading/trailing space
\hline\hline
Quantity & Meaning & Value  \\
\hline
\hline
$T$ & period of oscillation of parameters  &  $400$ \\
$h$ & Integration step  & $10^{-4}$ \\
$N$ & number of points in the ensemble & $N = 2 \times 10^4$ \\
$E_0$  & initial energy & $3 \times 10^5$\\
\hline
\end{tabular}
\label{tab1}
\end{table}
In the numerical simulations, we always start with initial conditions uniformly distributed at the energy level $E_0$.
Throughout the paper $\langle \cdot \rangle$ stands for the ensemble average with respect to these initial conditions.
All the relevant parameters for the simulations and their physical meaning can be found in Table \ref{tab1}.

\subsection{Symplectic integrator}

We implemented an explicit fourth-order method for the integration of equations of motion.
This method was used to eliminate non-symplectic effects while providing the accuracy of a
fourth-order integration step.  Our Hamiltonian is separable, meaning that it can be written in the form
$$
H(\bm{p},\bm{q}) = A(\bm{p}) + V(\bm{q}),
$$
with $A$ being the kinetic energy and $V$ the potential energy. Introducing the symbol
$
\bm{z} = (\bm{q},\bm{p})
$
and the operator,
$
D_H = \{ \cdot , H\},
$
which returns a Poisson bracket of the operand, the Hamilton's equation can be written
as $\dot{\bm{z}} = D_H \bm{z}$. The formal solution of this set of equations is given as
$\bm{z}(h) = \exp(h  D_H)\bm{z}(0)$. For this class of Hamiltonians the solution reads
$$
\bm{z}(h) = \exp[h  (D_A + D_V)]\bm{z}(0).
$$

The sympletic integration scheme approximates the time-evolution operator
$\exp[h ( D_A + D_V)]$ in the formal solution by a product of operators as
$$
\exp[h (D_A + D_V)] = \prod_{i=1}^k \exp(c_i h D_A) \exp(d_i h D_V) + O(h^{k+1})
$$
For the fourth order integrator \cite{Sym} we have
\begin{eqnarray}
c_1 &=& c_4 = \frac{1}{2(2-2^{1/3})} \, \mbox{  and  } \,  c_2 = c_3 = \frac{1-2^{1/3}}{2(2-2^{1/3})} \nonumber  \\
d_1 &=& d_3 = \frac{1}{2 - 2^{1/2}} \mbox{ ,  }\, d_2 = - \frac{2^{1/2}}{2 - 2^{1/3}} \mbox{  and  } d_4 = 0.
\nonumber  \\
\end{eqnarray}

As our Hamiltonian is non-autonomous, we
therefore we can introduce $p_t$, a canonically conjugate variable of $t$, so the extended Hamiltonian becomes
$$
{\cal H}(\bm{p},\bm{q},t) = A(\bm{p}) + p_t + V(\bm{q},t).
$$
This quantity must be conserved by the integration scheme up to the given accuracy.
We checked the accuracy both with the time-independent and time-dependent Hamiltonian and found excellent agreement with the order of the method, that is, the Hamiltonian
$\cal H$ conserved up to order $O(h^{5})$ per integration step. Thus, the changes of the energy we observe in the time-dependent Hamiltonian system are due to dynamics
and are not induced by the numerical scheme.

\subsection{Integrable Regime}

\begin{figure}[!ht]
\centering
\includegraphics[width=8.6cm]{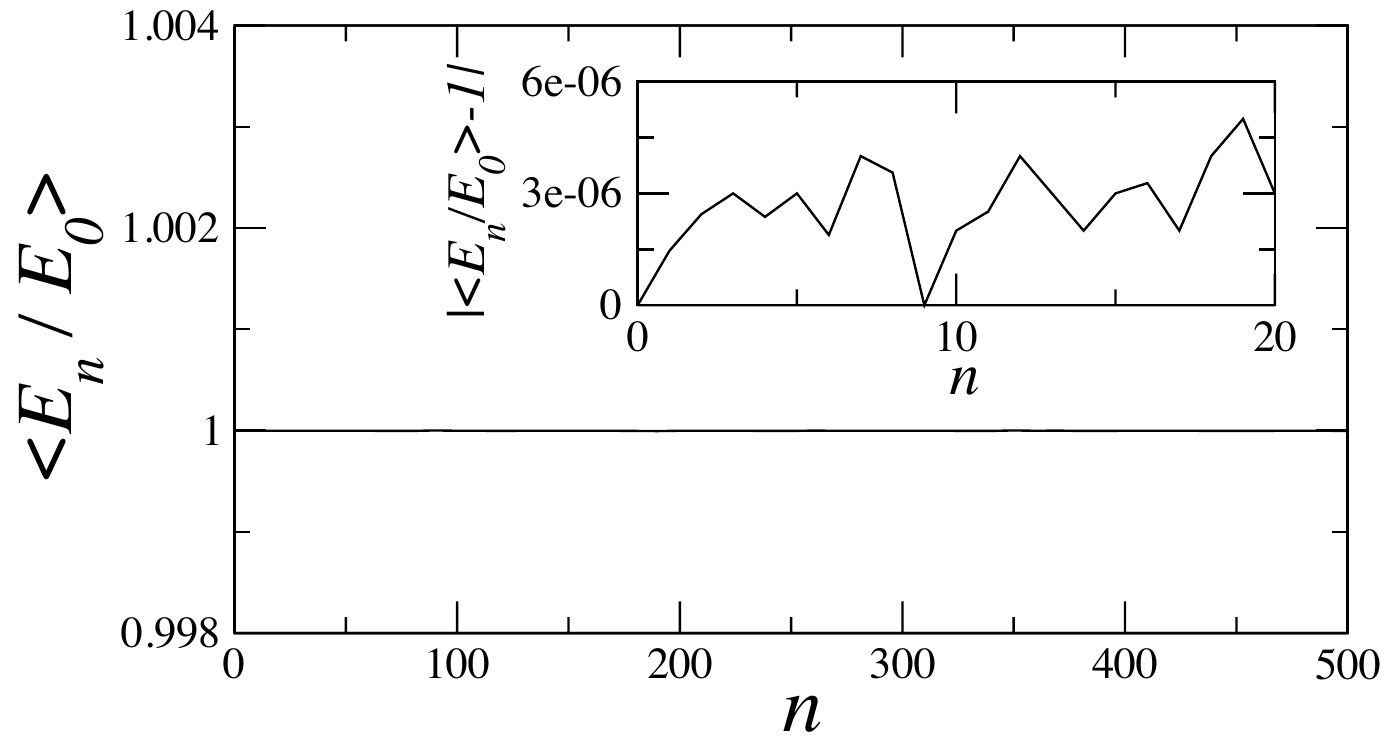}
\caption{{\bf Lack of energy growth in the integrable regime}.
We change parameters so that the frozen system is integrable for all values of the parameters. In these simulations we chose $A=1$. The ensemble averaged energy versus time is shown.}
\label{Integrable}
\end{figure}

If the parameters $a$ and $b$ are periodic functions of time, when keeping the Hamiltonian
within a given regime of integrability we have observed no energy growth.  For example, choosing the following parameters
$$
b = 0.0 \, \,   \mbox{    and    }   \, \, a = A \left[ 1.1 + \cos\left( \frac{2 \pi t}{T}\right) \right]
$$
the two systems are decoupled with $A>1$.  For example, choosing $A=1$ a typical behaviour of the energy can be seen in Figure \ref{Integrable}. We have experimented with $A$ ranging from $1$ to $50$ we found that no energy growth.
Another possibility is  $a = b$ depending periodically on time with period $T$.
Within this regime  we found no energy growth. For example, choosing
$b = A \times [1.1 +\cos(2\pi t/T)].$ We have experimented with $A$ ranging from $1$ to $50$
and found no energy growth.

\subsection{Strong chaos and polynomial energy growth}
\begin{figure}[!ht]
\centering
\includegraphics[width=9cm]{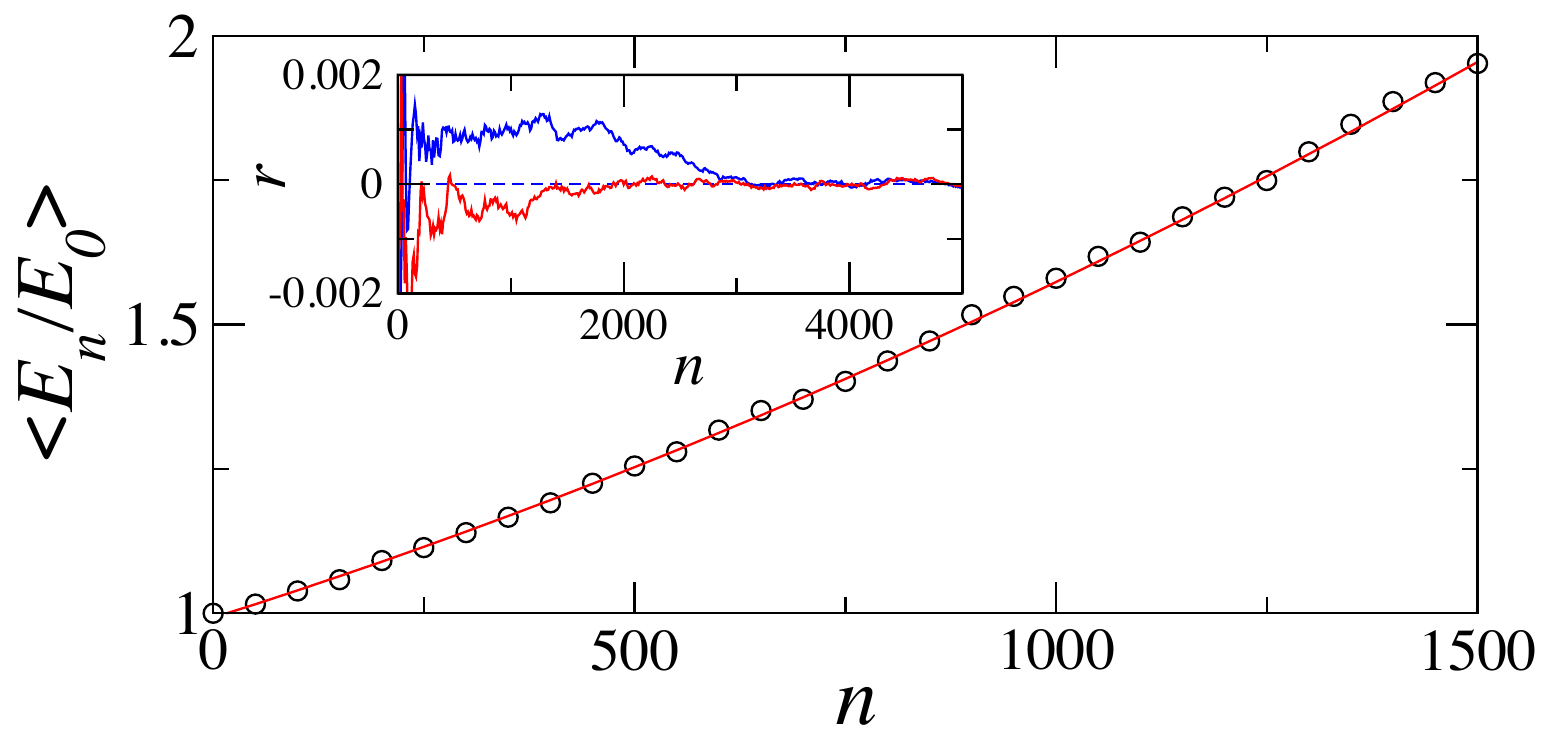}
\caption{{\bf Polynomial energy growth in the ergodic regime}.
We change parameters so that the frozen system remains chaotic, with no visible stability islands. The ensemble averaged energy versus time is shown.
The solid (red) line is the least-square fit by a quadratic polynomial.
In the inset the rates $r(n)=\frac{1}{n} \ln \left[E_n/E_0\right]$ are shown for two trajectories for a larger number of periods.
Both rates tend to zero, corroborating the lack of exponential acceleration.}
\label{ChaosOnly}
\end{figure}

If we change parameters in such a way that for each frozen value of the parameters the Hamiltonian remains strongly chaotic, we observe only
a slow energy growth. For instance, for the parameters
$$
a=0.01 \, \,  \mbox{  and   } \, \, \,   b(t) = 1.5 +\cos ( 2\pi t /T )
$$
the ensemble energy growth $\langle E_n/ E_0 \rangle$ versus the number of periods $n$ is shown in Fig. \ref{ChaosOnly}, which clearly reveals a quadratic growth
of the ensemble energy. For each initial condition, we also measure the energy gain after $n$ periods:
$$
\frac{E_n}{E_0} = \prod_{k=1}^{n-1} \frac{E_{k}}{E_{k-1}},
$$
and compute the exponential rates
$$
r(n) = \frac{1}{n} \ln \left[E_n/E_0\right].
$$
In the inset of Fig. \ref{ChaosOnly} we plot the exponential rate $r$ for two distinct initial conditions. As we see, $r$ converges to zero corroborating
the lack of exponential energy growth.

\subsection{Mixed phase Space and Exponential Acceleration}

\begin{figure*}[!ht]
\centering
\includegraphics[width=7cm]{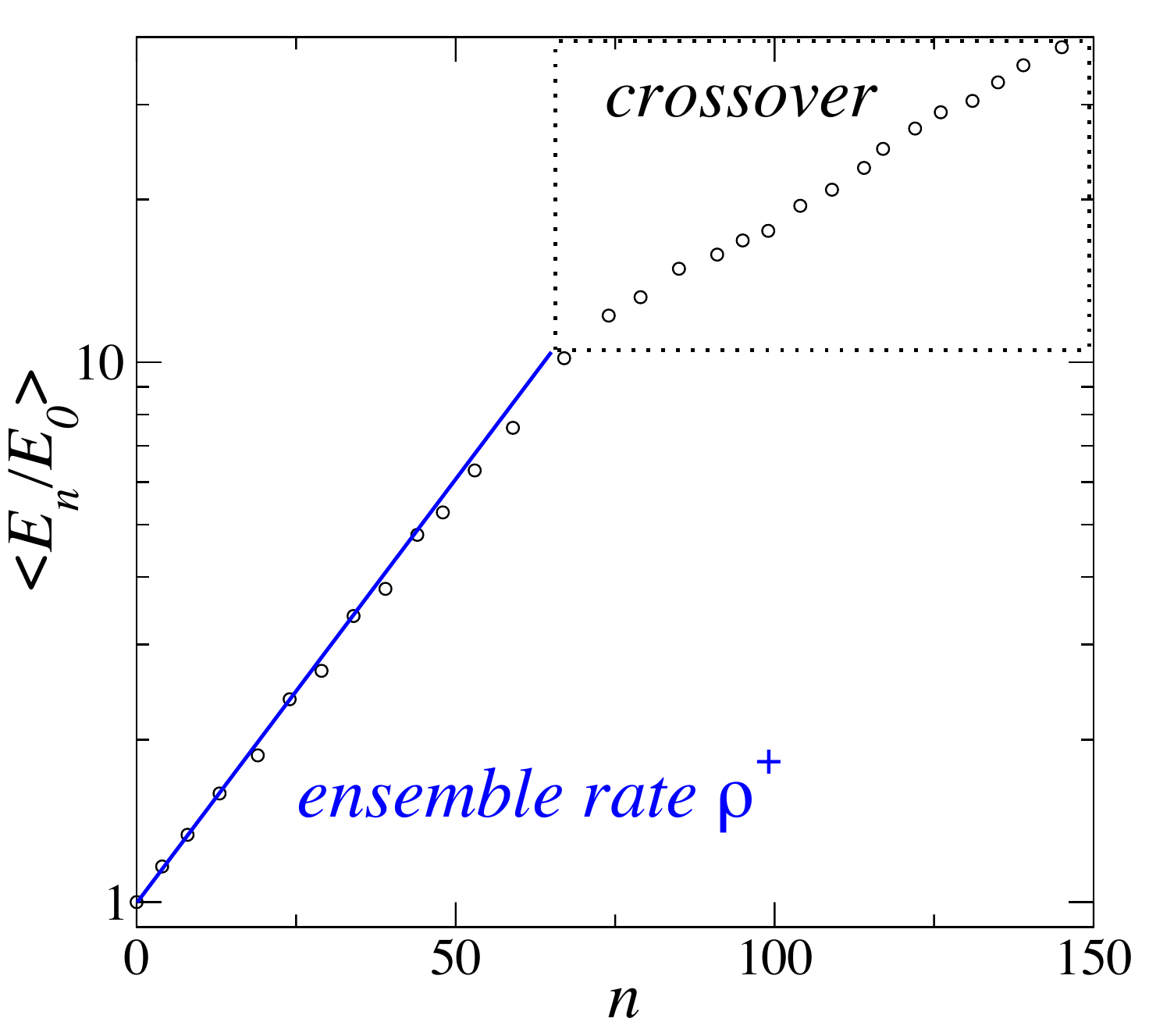}
\caption{{\bf Exponential energy growth}. We show the ensemble energy in log-scale versus the number of parameter cycles. The  parameters cycle described by equations \ref{Cya} and \ref{Cyb} for $a_0=0.1$, $A =1$, see Figure \ref{Fig_par}. For these parameters, the frozen Hamiltonian exhibits 
chaotic dynamics (along $a=0.01$), quasiperiodic motion (along $b=0$), and mixed behaviour (along the connecting arc). Repeating the cycle multiple
times, we observe exponential growth of energy. For the first $70$ cycles, we observe exponential growth at a higher ensemble rate $\rho^+$,
then a crossover to a lower rate starts. This crossover is a characteristic feature of a Geometric Brownian Motion.}
\label{Fig_ExpG}
\end{figure*}

\begin{figure*}[!ht]
\centering
\includegraphics[width=13cm]{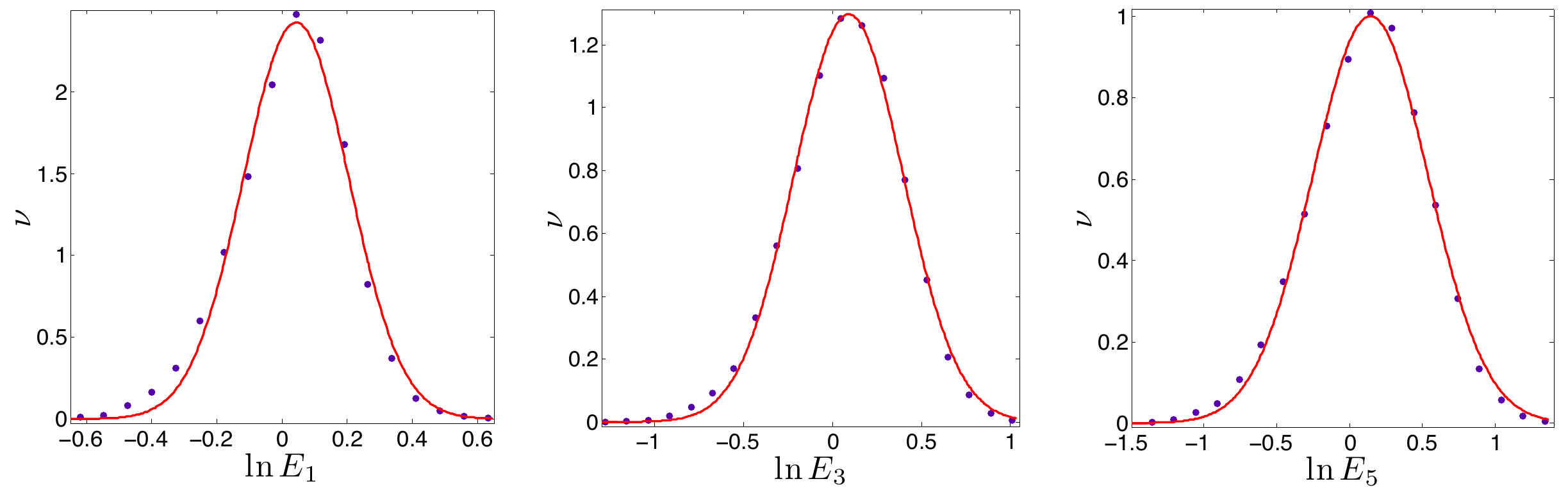}
\caption{{\bf Distribution of the logarithm of the energy}. Initially all particles start at the same energy level $E_0$, then after $n$ cycles of the parameters,
that is, after a time $nT$, we measure the logarithmic energy
gain $\ln ( E_n / E_0)$. The distribution of the energy gains $\nu$ is shown at $n=1$ in inset a), $n=3$ inset b), and $n=5$ for inset c).
The full line (red) shows the Gaussian fit.}
\label{Fig_EnD}
\end{figure*}

The most of our numerical experiments correspond to the case where the parameters visit both chaotic and integrable regions in the parameter space. We demonstrate that
in this case the acceleration process is in a good agreement with the Geometric Brownian Motion (GBM) model, as predicted by the theoretical arguments in
Sections \ref{ac} and \ref{entropy}.
The parameters are changed along the cycle displayed in Fig. \ref{Fig_par}, which is described by
\begin{equation}\label{Cya}
a =
\left\{
\begin{array}{cc}
A \cos\left( \frac{2\pi t}{ T}\right),  & \mbox{   if   } \cos\left( \frac{2\pi t}{ T}\right) >a_0 \\
a_0, & \mbox{  otherwise  }
\end{array}
\right.
\end{equation}
together with 	
\begin{equation}\label{Cyb}
b =
\left\{
\begin{array}{cc}
A \sin\left( \frac{2\pi t}{ T}\right),  & \mbox{   if   } \sin\left( \frac{2\pi t}{ T}\right) > 0 \\
0, & \mbox{  otherwise  }
\end{array}
\right.
\end{equation}
In Figs. \ref{Fig_ExpG} and \ref{Fig_EnD}, we show results for $a_0=0.1$, $A =1$.

The exponential energy growth is clearly seen in Fig. \ref{Fig_ExpG}. Note
that two distinct growth rates are observed. This is a characteristic signature of multiplicative random processes:
when the averaging is performed over a finite ensemble, process (\ref{claim1}) yields
$$
\mbox{ the rate} \, \, \, \rho^{+}=\ln \mathbb{E} \zeta \mbox{ for  small } n, \mbox{ then a crossover to the lower rate  } \, \,
\rho=\mathbb{E} \ln \zeta \, \, \, \mbox{ for large  } n
$$
(typically, the rate $\rho$ is established only after very large $n$; e.g. it was unachievable in our experiments, so we can see only the beginning of the cross-over
process). The cross-over phenomenon is well known \cite{Ole,RST,GRST}; it is explained by the fact that the variance ${\rm Var} (E_n)$ grows much faster than $\mathbb{E}(E_n)$.
In particular, when process (\ref{claim1}) is GBM (i.e., $\ln \zeta$ is normally distributed) the variance is given by
\begin{equation}\label{var}
\mbox{Var}(E_n) = e^{2 \rho^+ n } \left( e^{ \sigma^2 n} -1 \right)
\end{equation}
where
\begin{equation}\label{sig}
\sigma^2 = 2  ( \rho^+ - \rho)
\end{equation}

In Fig. \ref{Fig_EnD} we show the energy distribution after few periods of parameter oscillations (initial distribution
is a Dirac delta-function, i.e., we start with uniformly distributed initial condition on the energy shell $E_0$).
One can see that the logarithm of the energy $\ln E_n$ is well described by a Gaussian distribution,
already after the first cycle. This gives a qualitative agreement with our claim (\ref{distLogE}).

In order to make a quantitative check of our GBM model, we estimate the parameters of the GBM from the data.
The values of ensemble rate  $\rho^+$, single particle rate $\rho$ and the variance are estimated as
$$
\hat{\rho} = \frac{1}{n} \left\langle \ln \frac{E_n}{E_0}\right \rangle, \, \, \,
\hat{\rho}^+  = \frac 1 n \ln \left \langle \frac{E_n}{E_0} \right \rangle, \, \,\,
\mbox{  and   }  \, \, \, \hat{\theta}_n= \mbox{Var}(E_n/E_0).
$$
In our experiment, the values of the estimators stabilise already at the first cycle. Next, we estimate a relative error
with which the GBM predictions (\ref{var}),(\ref{sig}) hold:
$$
\chi = \frac{\left| \hat \theta^2_n -
e^{2 \hat{\rho}^+ n } \left( e^{2( \hat{\rho}^+ - \hat \rho) n} -1 \right) \right|}{\hat \theta^2}.
$$
In Table 2, we show the estimators quantities for the first five cycles. The estimators give excellent agreement with the GBM model,
with the relative error $\chi$ about 3\%.
\begin{table}[ht]
\caption{Value of the estimators for the GBM for parameters in  (\ref{Cya}) and (\ref{Cyb}) for $a_0 = 0.01$ and $A = 1$.}
\centering
\begin{tabular}{ c c c c c} % Column formatting, @{} suppresses leading/trailing space
\hline\hline
$n$ & $\hat{\rho}^+$ & $\hat \rho$ & $\hat \theta^2$  & $\chi$  \\
\hline
\hline
$1$ & $0.0391$ & $0.0256$ & $0.0305$  & $0.029$  \\
$3$ & $0.039$ & $0.0234$ & $0.1175$ & $0.055$ \\
$5$ & $0.0387$ & $0.0238$ & $0.2307$ & $0.025$\\
\hline
\end{tabular}
\label{tab2}
\end{table}

\begin{figure}[!ht]
\centering
\includegraphics[width=10cm]{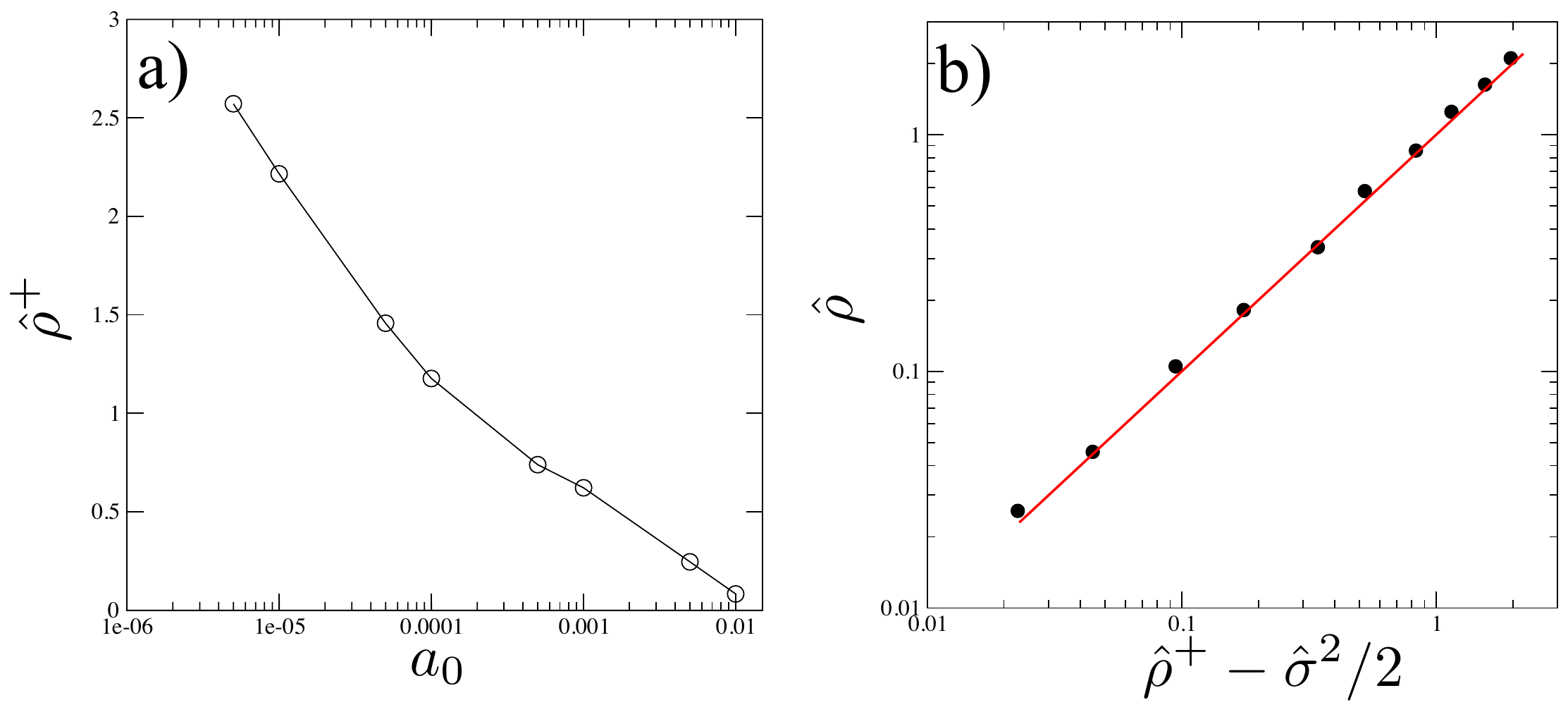}
\caption{{\bf Geometric Brownian motion for various parameter values}. Inset a) shows the estimated ensemble rate $\hat \rho^+$ versus
the parameter $a_0$.  For each value of $a_0$ we estimate the GBM parameters and check relation \ref{sig}. In inset b) we plot the estimate
$\hat \rho$ obtained from data against the parameters $\hat \rho^+$ and $\hat \sigma^2$ also obtained from the data. The full circle shows the
results whereas the full line (red) is the identity line. The results are in excellent agreement with the GBM model.}
\label{Fig_check}
\end{figure}

In order to see how sensitive is the GBM prediction to the variations of parameters of the system, we use the following procedure.
From the data, using the values obtained for $\hat \theta$, we estimate $\sigma^2$ by formula (\ref{var}).
Denote this estimated value $\hat \sigma^2$. Then we can systematically check the GBM relation (\ref{sig}), which is a consequence of (\ref{distLogE}).
To this end, we notice that by changing the parameter $a_0$ we can tune the ensemble rate $\rho^+$. In Fig.\ref{Fig_check}a we
vary $a_0$ in the interval $[10^{-6},10^{-2}]$ and compute the ensemble rate after one cycle of parameters oscillation. As $a_0$ decreases, the volume of the phase space
(of the frozen system) increases, enhancing the exponential rate $\rho^+$. Then, for each value of $a_0$ we compute $\rho^+$, $\rho$ and $\sigma$ and
verify relation (\ref{sig}). The result can be seen in Fig. \ref{Fig_check}b), which shows excellent agreement with the GBM model.

%\begin{acknowledgement}
{\bf Acknowledgement}: 
This paper is an extended version of the short note \cite{PT}.
We are grateful to V. Gelfreich and V. Rom-Kedar for useful discussions.
T.P. was supported by UK Leverhulme Trust Grant No. RPG-279 and European Union FP7 Marie Curie Project 303180.
D.T. was supported by Grant No. 14-41-00044 of RSF (Russia) and by the Royal Society grant IE141468.
%\end{acknowledgement}

%*******************************************************************
%BIBLIOGRAPHY
%*******************************************************************

%%%%%%%%%%%%%%%%%%%%%
%%%%%%%%%%%%%%%%%%%%%


\begin{thebibliography}{99}

\bibitem{Fermi} E. Fermi. {\it On the Origin of the Cosmic Radiation}. Phys. Rev. 75 (1949), 1169.

\bibitem{Ulam} S.M. Ulam, {\it On some statistical properties of dynamical systems}. Proc. 4th Berkeley Symp. Mathematical Statistics and Probability, Vol. 3, pp. 315-320 (University of California Press, Berkeley 1961).

\bibitem{Kochar} E. V. Derishev, V. V. Kocharovsky, Vl. V. Kocharovsky, {\it Cosmic accelerators for ultrahigh-energy particles}. Physics - Uspekhi 50 (2007),  308.
%  

%Citation: E. V. Derishev, V. V. Kocharovsky, V. V. Kocharovskii, “Cosmic accelerators for ultrahigh-energy particles”, UFN, 177:3 (2007), 323–330 

\bibitem{plasma} M. A. Lieberman and V. A. Godyak. {\it From Fermi acceleration to collisionless discharge heating}. IEEE Trans. Plasma Sci. 26 (1998), 955.
%From Fermi acceleration to collisionless discharge heating
%MA Lieberman, VA Godyak
%Plasma Science, IEEE Transactions on 26 (3), 955-986
%


\bibitem{nuc} J. Blocki et al., {\it One-body dissipation and the super-viscidity of nuclei}. Ann. Phys. (N.Y.) 113 (1978), 330.
% 

\bibitem{1dpeople} L.D. Pustyl'nikov. {\it On Ulam's problem}. Theor. Math. Phys. 57, 1035 (1983).
% L. D. Pustyl'nikov, “”, Teoret. Mat. Fiz., 57:1 (1983), 128–132 


\bibitem{P95} L. D. Pustyl'nikov. {\it Existence of Invariant curves for maps close to degenerate maps, and a solution of the Fermi-Ulam problem.} Sb. Math. 82 (1995), 231.
%EXISTENCE OF INVARIANT CURVES FOR MAPS CLOSE TO DEGENERATE MAPS, AND A SOLUTION OF THE FERMI-ULAM PROBLEM

\bibitem{LL} S. Laederich, M.Levi. {\it Invariant curves and time-dependent potentials}, Ergod. Th. \& Dynam. Sys. 11 (1991), 365.
%   Ergod. Th. & Dynam. Sys., 11 (1991), pp. 365–378.

\bibitem{Dov} A. Dovbysh. {\it The separatrix of an unstable position of equilibrium of a Hess-Appelrot gyroscope}. J. Appl. Math. Mech. 56 (1992), 188.
%  Volume 56, Issue 4, 1992, Pages 534–545


\bibitem{LRA1} A. Loskutov, A.B. Ryabov, L.G. Akinshin, {\it Mechanism of Fermi acceleration in dispersing billiards with time-dependent boundaries}. JETP 89 (1999), 966.
%Mechanism of Fermi acceleration in dispersing billiards with time-dependent boundaries
%Journal of Experimental and Theoretical Physics 89 (5), 966-974

\bibitem{LRA2} A. Loskutov, A.B. Ryabov, L.G. Akinshin. {\it Properties of some chaotic billiards with time-dependent boundaries}. J. Phys. A: Math. Gen. 33 (2000), 7973.
% 
%Journal of Physics A: Mathematical and General 33 (44), 7973 (2000)




\bibitem{2dpeople} A. Loskutov and A. Ryabov. {\it Particle dynamics in time-dependent stadium-like billiards}. J. Stat. Phys. 108 (2002), 995.
%
%Journal of Statistical Physics 108 (5), 995-1014

\bibitem{CSL1} R.E. de Carvalho, F.C. de Souza, E.D. Leonel. {\it Fermi acceleration on the annular billiard: a simplified version}. J. Phys. A: Math. Theor. 39 (2006), 3561.
%

\bibitem{CSL2} R.E. de Carvalho, F.C. de Souza, E.D. Leonel. {\it Fermi acceleration on the annular billiard}. Phys. Rev. E 73, 066229 (2006).
%

\bibitem{LDS} F. Lenz, F.K. Diakonos, and P. Schmelcher. {\it Tunable Fermi Acceleration in the Driven Elliptical Billiard}.  Phys. Rev. Lett. 100 (2008), 014103.
% Tunable Fermi Acceleration in the Driven Elliptical Billiard

\bibitem{LOL} E.D. Leonel, D.F.M. Oliveira, A. Loskutov. {\it Fermi acceleration and scaling properties of a time dependent oval billiard}. Chaos 19 (2009) 033142.
%

\bibitem{OVL}  D.F.M. Oliveira, J.Vollmer, E.D.Leonel. {\it Fermi acceleration and its suppression in a time-dependent Lorentz gas}. Physica D 240 (2011), 389.
%Fermi acceleration and its suppression in a time-dependent Lorentz gas

\bibitem{BR} B. Batisti\'c, M. Robnik. {\it Fermi acceleration in time-dependent billiards: theory of the velocity diffusion in conformally breathing fully chaotic billiards}. J. Phys. A: Math. Theor. 44 (2011), 365101.
%Semiempirical theory of level spacing distribution beyond the Berry–Robnik regime: modeling the localization and the tunneling effects


\bibitem{Jar} C. Jarzynski. {\it Energy diffusion in a chaotic adiabatic billiard gas}. Phys. Rev. E 48 (1993), 4340.
%

\bibitem{BunimovichLeonel} E.D. Leonel and L.A. Bunimovich. {\it Suppressing Fermi acceleration in a driven elliptical billiard}. Phys. Rev. Lett. {\bf 104} (2010), 224101.
%

\bibitem{OR} D.F.M. Oliveira, M.Robnik. {\it In flight dissipation as a mechamism to suppress Fermi acceleration}. Phys. Rev. E 83  (2011), 026202.
%

\bibitem{GRST} V.Gelfreich et al., {\it Robust exponential acceleration in time-dependent billiards}. Phys. Rev. Lett. 106 (2011), 074101.
%

\bibitem{GRT} V. Gelfreich, V. Rom-Kedar, and D. Turaev. {\it Fermi acceleration and adiabatic invariants for non-autonomous billiards}. Chaos 22 (2012), 033116.
%

\bibitem{An} D.V. Anosov, {\it Averaging in systems of ordinary differential equations with rapidly oscillating solutions}. Izv. Akad. Nauk SSSR, Ser. Mat. 24 (1960), 721.
% 

\bibitem{Kas} T. Kasuga. {\it On the adiabatic theorem for the Hamiltonian system of differential equations in the classical mechanics, I.} Proc. Jpn. Acad. 37  (1961), 366.
% 

\bibitem{Ott} E. Ott, {\it Goodness of Ergodic Adiabatic Invariants}. Phys. Rev. Lett. 42 (1979), 1628.
%

\bibitem{Gre} R. Brown, E. Ott, and C. Grebogi. {\it The Goodness of Ergodic Adiabatic Invariants}. J. Stat. Phys. 49, 511 (1987).
%

\bibitem{McK} R.S. MacKay,  Nonlinear Dynamics and Chaos: Advances and Perspectives, pp 89-102 (Springer 2010).

\bibitem{LoM} P. Lochak and C. Meunier, Multiphase Averaging for Classical Systems (Springer-Verlag, New York, 1988).

\bibitem{Herzentr} S. Hilbert, P. H\"anggi, J. Dunkel. {\it Thermodynamic laws in isolated systems}. Phys. Rev. E 90 (2014), 062116. 
%  

\bibitem{RST} K. Shah, D. Turaev, and V. Rom-Kedar. {\it Exponential energy growth in a Fermi accelerator}. Phys. Rev. E 81  (2010), 056205.
% ,

\bibitem{Kushal} K. Shah. {\it Energy growth rate in smoothly oscillating billiards}. Phys. Rev. E 83  (2011), 046215.
% 

\bibitem{mushroom} V. Gelfreich, V. Rom-Kedar, and D. Turaev. {\it Oscillating mushrooms: adiabatic theory for a non-ergodic system}. J. Phys. A 47  (2014), 395101.
%

\bibitem{bati} B. Batisti\'c. {\it Exponential Fermi acceleration in general time-dependent billiards}. Phys. Rev. E 90 (2014), 032909.
%

\bibitem{TRK} D.Turaev, V.Rom-Kedar. {\it Elliptic islands appearing in near-ergodic flows}. Nonlinearity 11 (1998), 575.
% ,

\bibitem{TV} D.Turaev, {\it Exponential Fermi acceleration in adiabatically perturbed Hamiltonian systems}. Proc. 8th European Nonlinear Dynamics Conference (ENOC 2014).
% Exponential Fermi acceleration in adiabatically perturbed Hamiltonian systems,

\bibitem{Ok} B.K. Oksendal,  {\it Stochastic Differential Equations: An Introduction with Applications}, Springer (2002).

\bibitem{Ole} O.Peters, W.Klein. {\it Ergodicity Breaking in Geometric Brownian Motion}.  Phys. Rev. Lett. 110 (2013), 100603.
%


\bibitem{Canergie} A. Canergie, I.C. Percival. {\it Regular and chaotic motion in some quartic potentials}. J. Phys. A 17 (1984), 801.
% 
%A Carnegie and I C Percival
%1984 J. Phys. A: Math. Gen. 17 801

\bibitem{Bon} M.V.S. Bonanca, M.A.M. de Aguiar. {\it Classical dissipation and asymptotic equilibrium via interaction with chaotic systems}.  Physica A 365 (2006), 333.
%

\bibitem{Sym} E. Forest, R.D. Ruth. {\it Fourth-order symplectic integration}. Physica D 43 (1990), 105.
%

\bibitem{PT} T.Pereira, D.Turaev. {\it Exponential energy growth in adiabatically changing Hamiltonian systems.} Phys. Rev. E 91 (2015), 010901(R).
 %

\end{thebibliography}
\end{document}